\documentclass[reprint,superscriptaddress,amsmath,amssymb,aps,pra,longbibliography]{revtex4-2}

\usepackage[colorlinks]{hyperref}
\usepackage{graphicx}
\usepackage{dcolumn}
\usepackage{bm}

\usepackage{siunitx}
\usepackage{bbm}
\usepackage{braket}
\usepackage{xcolor,verbatim}
\usepackage{mathtools}
\usepackage[normalem]{ulem}

\newcommand*\chem[1]{\ensuremath{\mathrm{#1}}}

\begin{document}

\title{Achievements and perspectives of optical Fiber Fabry-Perot Cavities}


\author{H. Pfeifer}
 \email{hannes.pfeifer@iap.uni-bonn.de}
 \affiliation{Institute of Applied Physics, University of Bonn,
              Wegelerstr. 8, 53115 Bonn, Germany}
\author{L. Ratschbacher}
 \affiliation{Institute of Applied Physics, University of Bonn,
              Wegelerstr. 8, 53115 Bonn, Germany}
 \affiliation{Intelligent Wireless Systems, Silicon Austria Labs, 
				Altenberger Strasse 66c, Linz, Austria}
\author{J. Gallego}
 \affiliation{Institute of Applied Physics, University of Bonn,
              Wegelerstr. 8, 53115 Bonn, Germany}
 \affiliation{current address: Fremont, CA, USA}
\author{C. Saavedra}
 \affiliation{Institute of Applied Physics, University of Bonn,
              Wegelerstr. 8, 53115 Bonn, Germany}
\affiliation{División de Ciencias e Ingenierias 
              Universidad de Guanajuato, Mexico}
\author{A. Faßbender}
 \affiliation{Institute of Physics, University of Bonn,
              Nussallee 12, 53115 Bonn, Germany}
\author{A. von Haaren}
 \affiliation{Institute of Physics, University of Bonn,
              Nussallee 12, 53115 Bonn, Germany}
\author{W. Alt}
 \affiliation{Institute of Applied Physics, University of Bonn,
              Wegelerstr. 8, 53115 Bonn, Germany}
\author{S. Hofferberth}
 \affiliation{Institute of Applied Physics, University of Bonn,
              Wegelerstr. 8, 53115 Bonn, Germany}
\author{M. Köhl}
 \affiliation{Institute of Physics, University of Bonn,
              Nussallee 12, 53115 Bonn, Germany}
\author{S. Linden}
 \affiliation{Institute of Physics, University of Bonn,
              Nussallee 12, 53115 Bonn, Germany}
\author{D. Meschede}
 \affiliation{Institute of Applied Physics, University of Bonn,
              Wegelerstr. 8, 53115 Bonn, Germany}
\date{\today}

\begin{abstract}
Fabry-Perot interferometers have stimulated numerous scientific and technical applications ranging from high resolution spectroscopy over metrology, optical filters, to interfaces of light and matter at the quantum limit and more. End facet machining of optical fibers has enabled the miniaturization of optical Fabry-Perot cavities. Integration with fiber wave guide technology allows for small yet open devices with favorable scaling properties including mechanical stability and compact mode geometry. These Fiber Fabry-Perot Cavities (FFPCs) are stimulating extended applications in many fields including  cavity quantum electrodynamics, optomechanics, sensing, nonlinear optics and more.

Here we summarize the state of the art of devices based on Fiber Fabry-Perot Cavities, provide an overview of applications and conclude with expected further research activities.
\end{abstract}


\maketitle

\section{Introduction}
\label{introduction}

For this interim review we summarize the state of the art of Fiber Fabry-Perot Cavities (FFPCs), i.e. Fabry-Perot resonators integrated on optical fibers. We discuss their basic properties and applications (with focus on gaseous media) and we conclude with an outlook into future perspectives. The article may also serve as a resource letter through its collection of references with relevance for FFPCs.

\subsection{Fabry-Perot interferometers}

Ever since their invention in the late 19$^\text{th}$ century Fabry-Perot interferometers (FPIs)~
\cite{fabryperot1899} have played an important role in advancing the resolution as well as the applications of optical spectrometers. In the simplest case they consist of two parallel plane and partially transmitting mirrors spaced by a length $\ell$. Multiple reflections of a light beam back and forth between its mirrors cause interference, where for normal incidence sequentially reflected beams have a path difference $2\ell$. Hence, constructive interference occurs when an integer number of wavelengths $\lambda$ covers the round trip path, $N\cdot \lambda = 2\ell$~\cite{bornwolf1959,siegman1986lasers}. The integer number $N$ is called the order of the interferometer and is typically large, of order $N \approx 10^3 - 10^6$. The path difference is cast into a resonance condition for the frequency of a light field ($\nu=c/\lambda$) by 
\begin{equation}\label{eq:fpires}
    \nu_N= N\cdot\Delta\nu_\text{FSR}\ ,
\end{equation}
where $\Delta\nu_\text{FSR}=c/2\ell$ is the free spectral range of the FPI with $\ell$ the cavity length. 
The width of an isolated resonance is determined by the reflectivities $R$ of the mirrors. A conveniently accessible measure for the spectral quality of the FPI is given by the ratio of the free spectral range and the Lorentzian line width $\Delta_\text{1/2}$ (FWHM) of a single spectral line, which is called finesse ${\cal F}$. In the simplest case of a symmetric cavity with two mirrors of identical reflectivity and negligible losses $A \ll 1{-}R$, it reads
\begin{equation}\label{eq:finesse}
    {\cal{F}} = \frac{\Delta_\text{FSR}}{\Delta_\text{1/2}}=\frac{\pi\sqrt{R}}{1-R}\ ,
\end{equation}
where for the frequent situation $R\to 1$ we have ${\cal{F}}\simeq \pi/(1-R) \gg 1$. 

With this, FPIs have offered the highest spectroscopic resolution in classic optical instrumentation before the advent of laser spectroscopy, acting essentially as narrow band optical filters with resolution~$\nu_N/\Delta_\text{1/2} = N\cdot{\cal{F}}$.

\subsection{Fabry-Perot cavities}

The light field circulates back and forth between the mirrors with a round trip time $\tau_\text{circ}=1/\Delta_\text{FSR}$. FPIs have therefore given birth to the notion of Fabry-Perot resonators or Fabry-Perot cavities (FPCs), where the focus is on the properties of the internally stored light field rather than the transmitted or reflected intensity. An initially stored light field is attenuated as a consequence of the finite mirror transmissivity $(1{-}R)$ and other losses $A$. For small losses and transmission, the relaxation rate $1/\tau_\text{cav}$ per round trip is approximately given by $1/\tau_\text{cav}=(1{-}R)/\tau_\text{circ}$ which translates into the mean number of round trips
    \begin{equation}\label{eq:circ}
    n_\text{round} = \frac{\tau_\text{cav}}{\tau_\text{circ}} \simeq \frac{{\cal{F}}}{\pi}\ .
    \end{equation}
The decay constant $\kappa$ of the amplitude of of the stored light field (energy relaxation rate $2\kappa$) is the relevant rate describing dynamic properties of any FPC. A spectrum of an FPC shows a Lorentzian line width (FWHM = $\Delta_\text{1/2}$), which is related to the relaxation rate by $\Delta_\text{1/2} = \kappa/\pi$.
FPCs have strongly inspired the advent of the laser -- an FPI/FPC containing an amplifier for optical waves -- rendering one of the most important tools in contemporary science and technology. 

The geometry of the coherent light fields provided by lasers are usually described in terms of Gaussian field modes, which are extensively covered in textbooks such as~\cite{siegman1986lasers}. 
\begin{figure}
    \centering
    \includegraphics[width=\columnwidth]{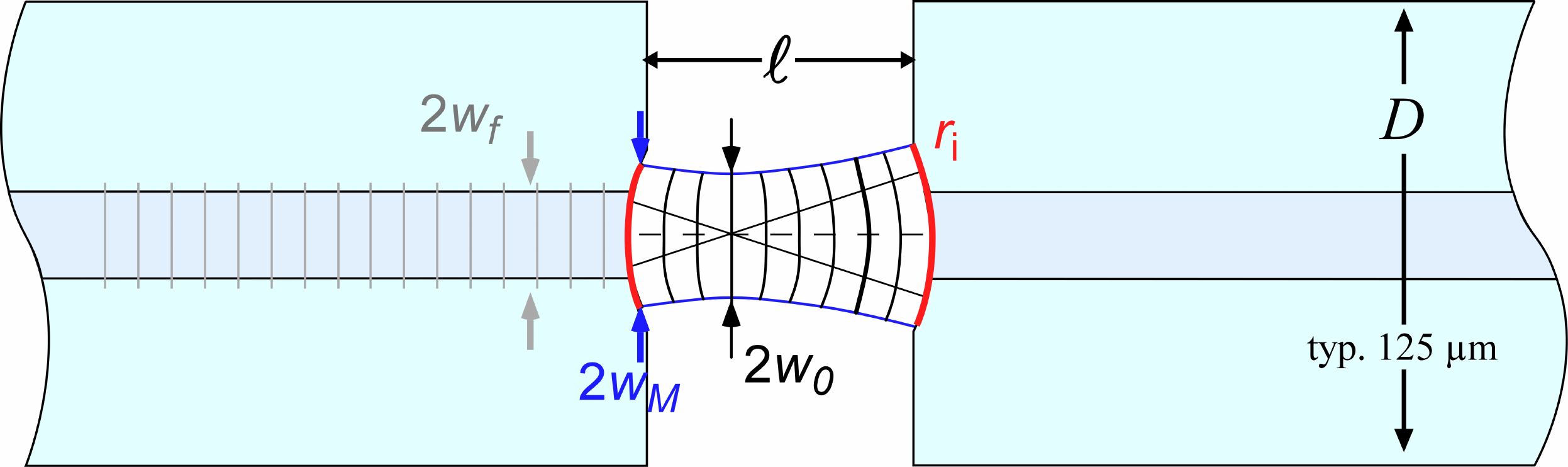}
    \caption{Basic geometric parameters of a TEM$_\text{00}$ mode supported by an optical fiber Fabry-Perot cavity (FFPC). Wavefronts of the cavity mode field match the radii of curvature of the integrated mirrors. The coupling strength with incoming light fields is determined by the overlap of the travelling guided and the cavity modes.}
    \label{fig:cavmode}
\end{figure}
The most widely used Gaussian mode is the so-called TEM$_\text{00}$, which is schematically shown in Fig.~\ref{fig:cavmode}. The properties of the fundamental cavity modes can be expressed in terms of the dimensionless cavity parameters $G_i$ (mirrors $i=1,2$), which are related to the mirror geometry and the cavity length $\ell$,
    \begin{equation}
        G_i = 1-\ell/r_i.
    \end{equation}
 The stability condition binds the individual $G_i$ values according to $0 \leq G_1G_2 \leq 1$ (i.e. eigensolutions of the paraxial Helmholtz equation, see~\cite{siegman1986lasers}), with special cases $G_1 = G_2 = 1$ (plane-plane cavity), 0 (confocal) and -1 (concentric).
 A stable cavity mode shows wave fronts matching the curvature of the mirrors with radii $r_i$. Most FPCs are arranged to have the focus, the position of highest field strength, within its volume. The waist is conveniently calculated from
    \begin{equation}\label{eq:w0g1g2}
         w_0^2 = \frac{\ell \lambda}{\pi}\sqrt{\frac{G_1 G_2(1-G_1G_2)}{(G_1+G_2-2G_1G_2)^2}}\,, 
    \end{equation} 
and the resonator mode size at the position of mirror M$_1$ is given by
    \begin{equation}\label{eq:wmg1g2}
         w^2_{M_1} = \frac{\ell \lambda}{\pi}\sqrt{\frac{G_2}{(G_1(1-G_1G_2)}}\,, 
    \end{equation} 
where $\lambda=c/\nu_\text{cav}$ is the resonant wavelength of the cavity. To illustrate physical properties, the symmetric case is sufficient in most cases. With $G=G_1=G_2$ Eq.~\ref{eq:w0g1g2} reduces to
    \begin{equation}\label{eq:w0}
        w_0^2 = \frac{\ell \lambda}{2\pi}\sqrt{\frac{1+G}{1-G}} = \frac{\ell \lambda}{2\pi}\sqrt{\frac{r}{\ell/2}-1}\,,
    \end{equation}
where it is easy to see that the mirrors of the FPC have to fulfill the condition $r>\ell/2$ for their radii of curvature. The smallest waist radii $w_0$ are obtained in the limit $r\to \ell/2$. In this case of the concentric cavity, however, the paraxial approximation of Gaussian beams is no longer valid, and diffraction limits the minimal waist to a radius of order $w_0 \sim \lambda/2$. For the most relevant Gaussian TEM$_\text{00}$ mode of the FPC, the resonance condition Eq.\ref{eq:fpires} is slightly modified by the so-called Guoy shift~\cite{siegman1986lasers},
 \begin{equation}\label{eq:HGModes}
\nu_{N} = \frac{c}{2\ell} \left(N+\frac{\arccos{(G)}}{\pi}\right).
\end{equation}
For all practical purposes discussed here, this small correction ($\arccos{G}/\pi \ll N$) is included in an effective cavity length $\ell = \ell_\text{eff}$, which also accounts for the pene\-tration of the cavity field into the stack of dielectric layers of the mirror coating.

FPCs are frequently used to maximize light matter coupling, and hence atoms, membranes, or spectroscopic samples are positioned at the waist of the cavity mode with diameter $2w_0$. The field strength of the cavity field can be considered a function of the total energy $U$ stored in the FPC. An important measure for the ratio of the stored electromagnetic energy to the maximum field strength is given by the mode volume $V_\text{mode}$. It is calculated by spatially integrating the field distribution inside the cavity normalized to the maximum field strength $\vec{{\cal{E}}}(w_0)$, resulting in
    \begin{equation}\label{eq:vmode}
        V_\text{mode} = \frac{\pi}{4}w_0^2\ell = V_\lambda \cdot 
        \left(\frac{\ell}{\lambda} \right)^2\left(\frac{r}{\ell/2}-1\right)^{1/2},\
    \end{equation}
with minimal mode volume $V_\lambda=(\lambda/2)^3$ for a cubic cavity resonant at wavelength $\lambda$. The total electromagnetic energy $U$ stored in the FPC relates to the maximum field amplitude  $|\vec{{\cal{E}}}_\text{max}|$ usually at the waist through
    \begin{equation}\label{eq:uvse}
        U = \frac{1}{2}\epsilon_0c^2\int_{V_\text{cav}} \text{d}V\, |\vec{{\cal{E}}}(\vec{r})|^2 = \frac{1}{2}\epsilon_0c^2|\vec{{\cal{E}}}_\text{max}|^2\cdot V_\text{mode}.
    \end{equation}

\section{Fiber Fabry-Perot cavities}
One route of the present evolution of photonics is the miniaturization of optical devices. In the past FPCs have successfully been integrated into optical ﬁbers through the inscription of Bragg mirrors directly into the glass fiber material, serving e g. as optical ﬁlters. The Fiber Fabry-Perot Cavities (FFPCs) introduced by~\cite{hunger2010fiber} and discussed here are intrinsically fiber connected, too. However, as scaled-down versions of classic FPCs with mirrors spaced by an empty volume, they enable versatile access to the ﬁeld mode stored in the cavity, which allows applications as outlined in Sec.~\ref{sec:applications}.

While the formal treatment of FFPCs is not different from the conventional description of macroscopic FPCs outlined above in terms of Gaussian beams and field modes, it is the limit of small typical length scales $\ell \sim 10- \SI{1000}{\micro\meter}$, which makes FFPCs a special limiting case. 

\begin{figure}[h]
    \centering
    \includegraphics[width=\columnwidth]{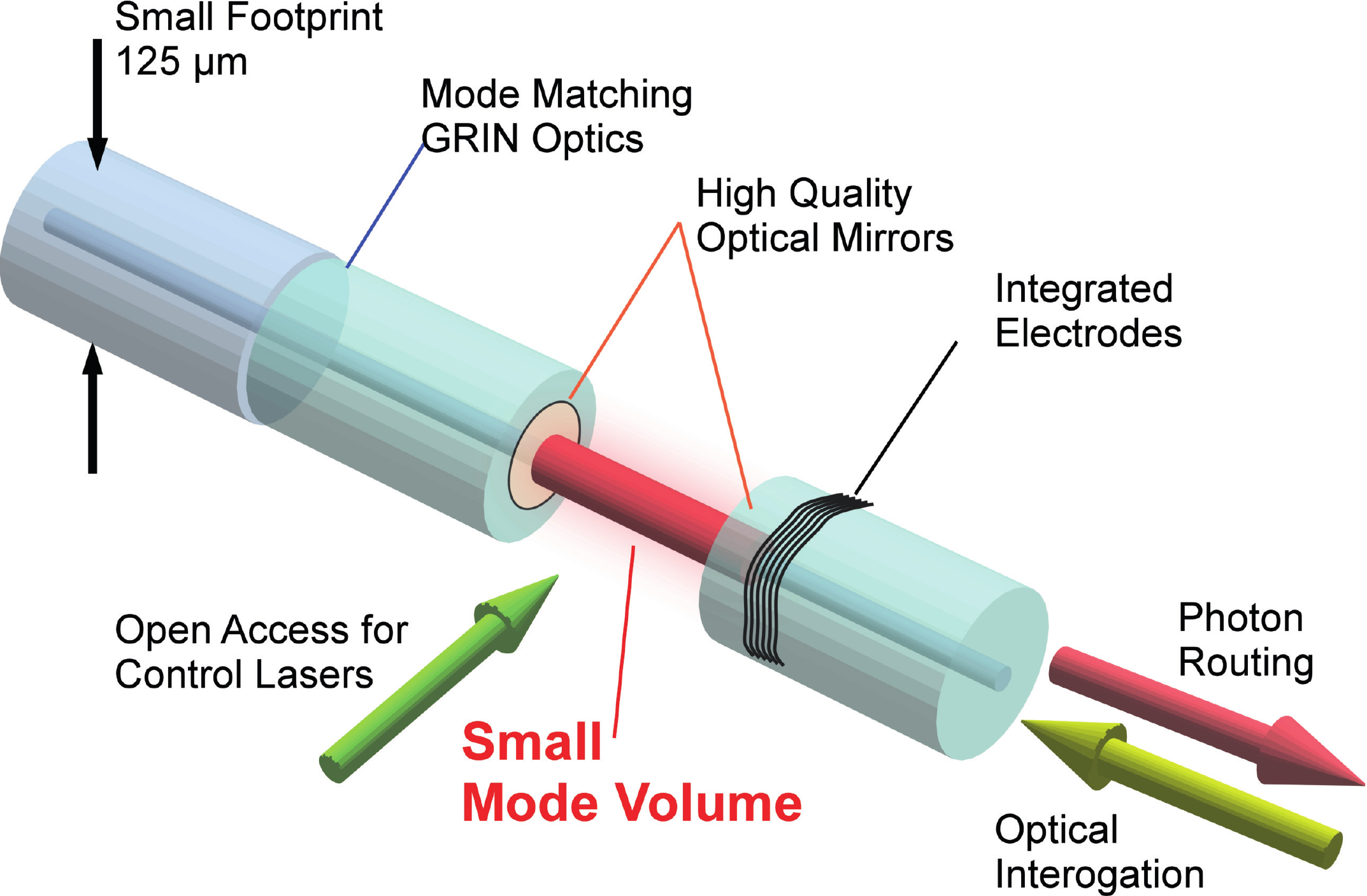}
    \caption{Concept of a Fiber Fabry-Perot Resonator (FFPC) integrated with optical fiber links and functional components.}
    \label{fig:ffpcfunc}
\end{figure}

A breakthrough of FFPCs was achieved when D. Hunger, J. Reichel and colleagues~\cite{hunger2010fiber,Steinmetz2006} developed the integration of mirrors with high optical quality on the end facets of optical fibers. With their continued work on FFPCs, it became clear that FFPCs offer several technical advantages over macroscopic devices due to their compactness and robustness (Fig.~\ref{fig:ffpcfunc}), including:
\begin{itemize}
    \item High field concentration
    \item Integration with optical fibers
    \item High optical quality
    \item Small foot print
    \item Open geometry
    \item Integration with other functional components
\end{itemize}
The following sections will briefly summarize the physical quantities associated with these aspects.

\subsection{FFPCs for high field concentration} \label{sec:fieldconc}

The realization of high local field strengths is at the heart of many FFPC applications. For all Gaussian modes sustained by FFPCs the TEM$_\text{00}$-mode offers the highest fields at its waist with diameter $2w_0$. According to Eqs.~\ref{eq:vmode} and \ref{eq:uvse} short cavity lengths $\ell$ and $\ell/r \to 2$ (cf. Eq.~\ref{eq:w0}) approaching the concentric cavity case let the mode volume decrease and the local field strength ${\cal{E}}_\text{cav}$ increase.

The global geometric scale of FFPCs for the relevant parameters like the cavity length $\ell$ and the radii of mirror curvatures $r_i$ is set by the standard diameter of optical fibers of $\SI{125}{\micro\meter}$ ($D$ in Fig.~\ref{fig:cavmode}). Choosing a small waist $w_0$ for high field concentration also means high divergence of the field mode. The divergence angle of a Gaussian mode is $\Theta_\text{div}=\lambda/(\pi w_0)$, and hence in order to avoid clipping losses by the fiber end facets we need to make sure that $D/2 > \Theta_\text{div}\cdot(\ell/2)$ yielding
    \begin{equation}\label{eq:lmax}
    \ell < \pi w_0 D/\lambda\ ,        
    \end{equation}
in agreement with the Rayleigh focusing limit for an aperture of diameter $D$. We use Eq.~\ref{eq:vmode} to estimate the range of mode volumes available for FFPCs with
    \begin{equation}
        V_\text{mode}/V_\lambda = \frac{\pi}{4}\left(\frac{w_0}{\lambda/2}\right)^3\left(\frac{\ell/2}{w_0}\right)
    \end{equation}
For realistic focusing conditions $5 \leq 2w_0/\lambda \leq 50$ we therefore obtain a wide range of possible mode volumes ranging from $V_\text{mode}/V_\lambda \sim 10^2 \cdot (2\ell/w_0)$ to $\sim 10^5 \cdot (2\ell/w_0)$, where Eq.~\ref{eq:lmax} sets an upper limit. All experimental realizations of FFPCs indeed fall into this regime, where the factor $2\ell/w_0$ may approach values below 10 when favorable small mode volumina are of interest.

As the mode volume reduces with the radii of curvature of the mirrors, high field concentrations are reached for small $r$ that, however, still have to fulfill the condition $r>\ell/2$. For strongly focusing mirrors the radii $r$ will, therefore, typically be on the order of $\ell$, i.e. in the range of some 10 to $\SI{1000}{\micro\meter}$.

\subsection{Mode matching for FFPCs}

Successful use of FFPCs implies efficient coupling of the cavity mode to the waveguide modes propagating within the fibers. Assuming perfect collinear alignment of the cavity axis and fiber axis the coupling efficiency into the cavity can be written as~\cite{hunger2010fiber}
    \begin{equation}
          \epsilon_\text{TEM00} = \frac{4}{(\frac{w_\text{f}}{w_\text{m}}+\frac{w_\text{m}}{w_\text{f}})^2+(\frac{\pi n_\text{f} w_\text{f}w_\text{m}}{\lambda r})^2}
    \end{equation}
with $w_\text{m}$, $w_\text{f}$, $n_\text{f}$ the waist of the cavity mode and of the fiber mode at the mirror, respectively, and the index of refraction of the fiber~\cite{hunger2010fiber}. The mode field waist of a standard single mode fiber, for instance, allows for a maximum coupling value of $\epsilon \approx 0.78$, considering a cavity for $\lambda = \SI{780}{\nano\meter}$ with $\ell = \SI{50}{\micro\meter}$, $r = r_1 = r_2 = \SI{150}{\micro\meter}$ and thus $w_0 \approx \SI{3.7}{\micro\meter}$ $w_{m}= \SI{4.1}{\micro\meter}$, as well as $w_{f}=\SI{2.53}{\micro\meter}$ and $n_{f}=1.46$. The mismatch in mode geometry of the incoming guided mode and the mode sustained by the FFPC can be mitigated by stacking fibers of different types to achieve mode shaping inside the fiber (cf. Sec.~\ref{FFPCdevices} on GRIN fibers). 

In practice, imperfect centering of the mirror curvature on the fiber center during fabrication and/or cavity misalignment lead to degraded coupling efficiencies. Detailed mode overlap calculations for misaligned cavity and fiber modes have been performed in Refs.~\cite{gallego2016high,Bick2016,gulati2017fiber} providing an explanation of asymmetric cavity line shapes as pure interference effects. The interplay of direct single mode fiber and cavity mode coupling results in alignment-dependent cavity line shapes in the reflected side at the incoupling fiber that are well described by the superposition of a Lorentzian and a dispersive component~\cite{gallego2016high}.        

\subsection{Finesse of FFPCs}
In addition to determining the fiber to cavity coupling, the alignment-dependent cavity mode properties also influence the FFPC finesse due to mirror clipping losses, cf. Eq.~\ref{eq:lmax}. The reduction in cavity finesse for longer cavities prior to the stability limit has been well explained by considering the cavity mode footprint at the position of the fiber mirrors to be clipped by mirrors with effective finite diameters~\cite{hunger2010fiber}. 

More recently, sharp variations in finesse at specific cavity lengths have been well modelled in position and magnitude as mirror-profile-dependent crossings of the fundamental cavity mode with higher order transversal modes. Deviations of the cavity mirrors from the idealized spherical mirror surface have also been identified as the origin of the birefringence, which are observed in many fiber cavity systems~\cite{benedikter2015transverse,podoliak2017harnessing,benedikter2019transverse}. 

\subsection{Polarization properties and birefringence}

The dominant contribution for frequency splitting of polarization eigenmodes in fiber Fabry–Perot cavities has been shown to result from the elliptic shapes of cavity mirrors forming the FPC. The lifting of the degeneracy of polarization modes expected from scalar field theory (cf. Eq.~\ref{eq:HGModes}) can be explained by extensions to vector theory that allows for field components along the cavity axis~\cite{Uphoff2015,garcia2018dual}. For many applications cavities with low birefringence are desirable and mirror manufacturing techniques (cf.~Sec.~\ref{fabrication}), including e.g. the rotation of the fiber during multi-shot sequences~\cite{takahashi2014}, have been introduced for this purpose. 

Recently, however, it was also shown that the birefringence of optical cavities can be turned into an advantage by making more field modes available for controlled light matter interaction~\cite{barrett2019,barrett2020}, e.g for further enhancing the Purcell effect, see Sec.~\ref{ssec:cqed}. 

\subsection{Fabrication of high quality mirrors for FFPCs}
\label{fabrication}

The small-radii of curvature required for cavities with small mode volumes are not within the reach of traditional polishing techniques. Successful techniques to fabricate the required small, smooth depressions (depth of order a few \textmu m) on glass surfaces are surface etching methods and laser ablation. Wet or ion-based etching has proven to be a useful tool for providing radii of curvature below $\SI{5}{\micro\meter}$, but with surface roughness at the nanometer scale~\cite{laliotis2012icppolish,trichet2015,Qing2019}, which cause limits in the achievable mirror reflectivities.

\begin{figure}[h]
    \centering
    \includegraphics[width=\columnwidth]{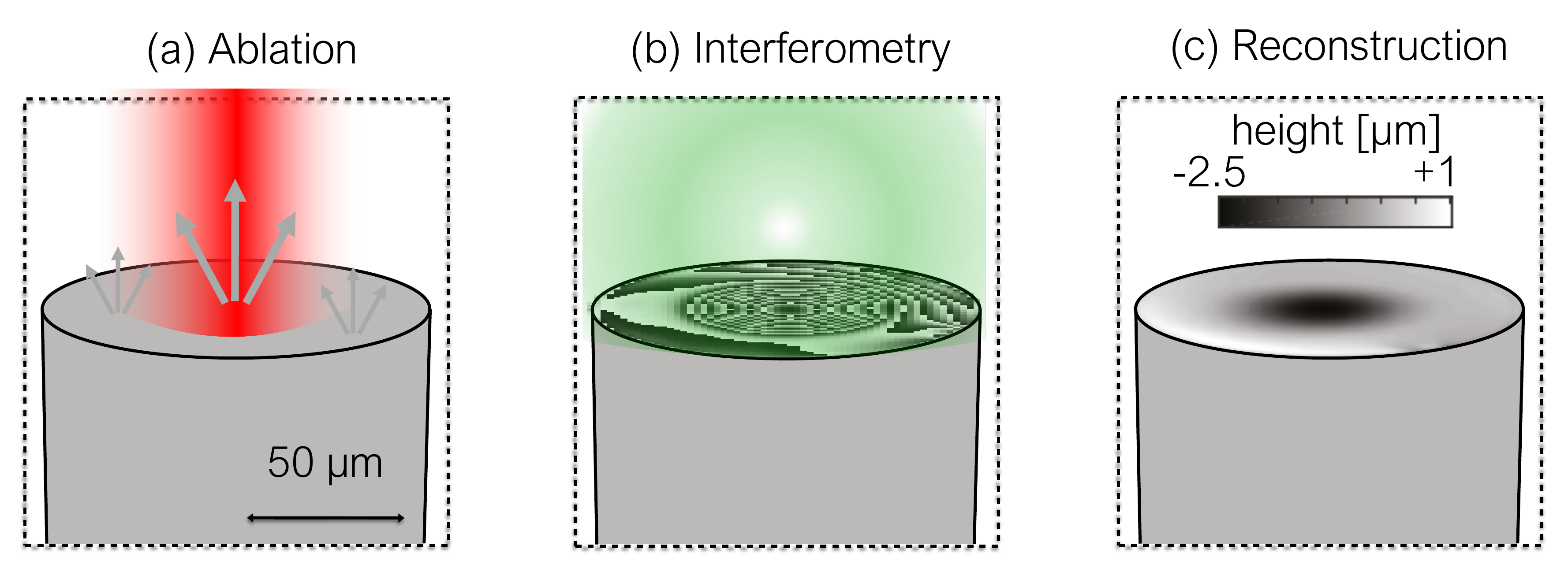}
    \caption{Treatment of fiber end facets for mirror fabrication. (a) Thermal ablation by pulsed CO$_2$ laser illumination. (b) Phase shift interferometry of the fiber end facet for quality control. (c) Numerical reconstruction of the fiber end facet surface through phase unwrapping.}
    \label{fig:fabrication}
\end{figure}

The most successful technique to date for producing curved mirror shapes with ultra-low surface roughness is \chem{CO_2}-Laser ablation (Fig.~\ref{fig:fabrication} (a)) ~\cite{hunger2010fiber},~\cite{muller2010lowmode},~\cite{hunger2012micro}. The strong absorption of fused silica (\chem{SiO_2}) for light between $9-\SI{11}{\micro\meter}$ wavelength is used to strongly heat the top layer of the fiber end surface by illumination with a focused \chem{CO_2} laser beam. For suitable intensity, focus and pulse duration the Gaussian laser intensity profile melts and evaporates part of the glass surface and produces a depression with a similar profile~\cite{hunger2010fiber,brandstaetter2013,takahashi2014}. The spherical part at the bottom of the resulting structure exhibits a low roughness comparable to those of superpolished mirrors.

In order to characterize the ablation process and to ensure the desired shape and alignment of manufactured depressions, the fiber mirrors are optically inspected. A typical setup uses a high numerical aperture microscope to image an LED illuminated fiber end in a Michelson Interferometry configuration to obtain interferogram images of the fiber surfaces with high spatial resolution (see Fig.~\ref{fig:fabrication} (b) and (c)).  

The ablation technologies have proven to be suitable for producing a wide range of fiber end microstructures, from micrometer-long cavities~\cite{kaupp2016smallmode} to elaborate geometric shooting patterns that allow the creation of fiber-resonators in the millimeter range~\cite{ott2016} and customized mirror ellipticities for tailored cavity birefringence~\cite{garcia2018dual}.  

Subsequently to the mirror surface molding, multilayer dielectric coatings are applied to realize low-loss, high-reflectivity mirrors in the visible, infrared or ultraviolet domain. For this purpose up to tens of alternate layers of materials with high (\chem{Ta_2O_5} (n=2.10) and low (\chem{SiO_2} n=1.45) index of refraction and individual thicknesses of $\lambda$/4 are deposited on the fiber surface using ion-beam sputtering techniques~\cite{muller2010lowmode}. The resulting fiber-based mirrors with the shape of the original fiber surface depression can feature transmission, scattering and absorption losses (after annealing~\cite{atanassova1995}) down into the few parts per million range, respectively and are ready to be aligned to build high-finesse, low-mode-volume resonators.

\subsection{FFPC devices}
\label{FFPCdevices}

Depending on the specific application, the geometries of FFPC implementations can strongly vary resulting in a large range of reachable specifications with regard to cavity and mode sizes, mode properties, degree of accessibility and stability of the devices.

\begin{figure}[h]
    \centering
    \includegraphics[width=\columnwidth]{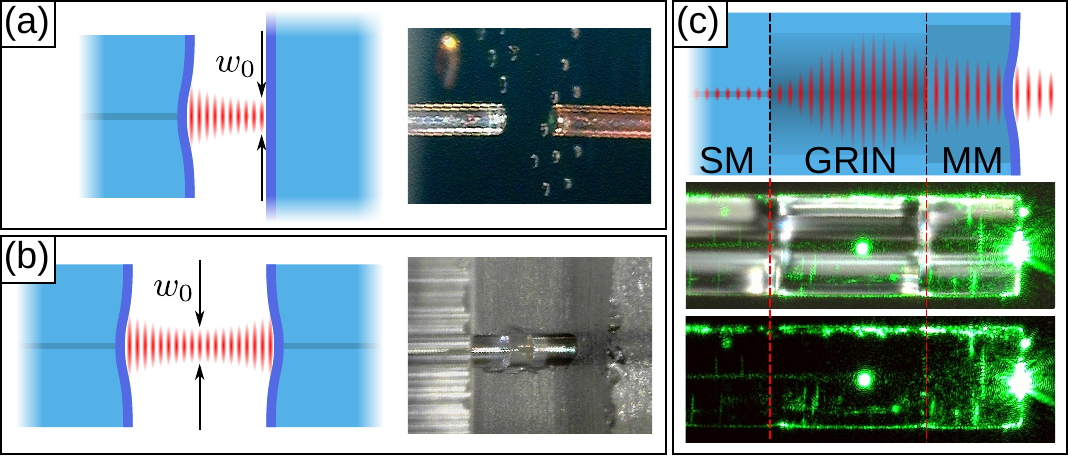}
    \caption{(a) shows a hemi-cavity, where the smallest mode waist is located on a flat mirror substrate. The amber colored fiber in the image is the reflection of the fiber mirror approached from the left. (b) in contrast is a symmetric cavity with the smallest mode waist located in the center. The first is used to interface structures on the surface of the flat mirror, while the latter is used to interact with gases, atoms or ions that can be trapped in the center of the cavity mode volume. (c) shows a schematic of mode-matching optics using a graded index fiber (GRIN) piece. The stack of different fiber pieces allows to shape the mode incident on the mirror from the fiber side and match it to the mode of the cavity~\cite{gulati2017fiber}.}
    \label{fig:fiberRealizations}
\end{figure}

FFPC devices can consist of either two opposing fiber mirrors or a single fiber mirror together with a usually flat, macroscopic mirror substrate~\cite{Steinmetz2006}. As shown in Fig.~\ref{fig:fiberRealizations}~(a)~and~(b), the implementations differ with respect to the position of the smallest cavity waist that is either located between the fiber mirrors or on the macroscopic mirror. Solid state quantum emitters on substrates are interfaced using the latter cavity geometry whereas experiments with trapped atoms or ions usually employ a symmetric cavity geometry in order to maximize the coupling to the emitter.

As many applications benefit from high $Q$ optical resonators, the finesse $\mathcal{F} = Q/N$ is one of the key specifications for FFPCs. In accordance with Eq.~\ref{eq:finesse}, it is determined by the losses $\mathcal{A}$ during a cavity round trip, with $\mathcal{F}=2\pi/\sum_i \mathcal{A}_i$. In FFPCs, the reachable finesse is usually limited by the quality of the reflective coating, the surface quality and the size of the fiber mirrors. Absorption in ion beam sputtered layer systems with high reflectivity usually occurs on the order of some parts per million. Annealing of these mirrors at moderate temperatures ($\leq \SI{350}{\degree}  $) can reduce the amount of absorption even further. Mirror carving using laser ablation creates smooth surfaces, but is hard to control close to the rim of the facet reducing the usable mirror size. Small effective mirror diameters however lead to clipping losses that increase with the separation of the fiber mirrors and reduce the finesse. Record finesses for FFPCs on the order of some 100000 have been reported~\cite{hunger2010fiber,rochau2021dynamical}.

The cavity length $\ell$ of FFPC devices can be customized depending on the target specification for mode volume and linewidth (cf. Sec.~\ref{sec:fieldconc}). Furthermore, it is a key parameter to set the resonator's free spectral range $\Delta\nu_\text{FSR}$ that requires consideration, if multiple cavity resonances are employed. The realization of specific cavity lengths requires the fabrication of tailored fiber mirrors with respect to the usable mirror size and the radius of curvature of the employed fiber mirrors. Record long FFPCs reach lengths on the order of \SI{1}{\milli\meter}~\cite{ott2016}. The length of the cavity is thereby limited by the size of the mode on the mirror with respect to the usable area, the ultimate limit being the size of the optical fiber (cf. Eq.~\ref{eq:lmax}). For mirrors fabricated using laser ablation the usable mirror size is usually much smaller than the full fiber facet. Mode waists larger than the spherical part of the mirror lead to clipping losses and hence strongly reduced finesse values. 

Another challenge especially for long FFPCs is mode matching of the fiber-guided and the cavity mode at the in-coupling mirror as the cavity mode size is increasingly large compared to usual fiber guided mode field diameters. For extremely short FFPCs with small mirror radii of curvature the mismatch of the wavefront curvature leads to a similar effect. To compensate for this effect in large cavities either fibers with a large mode-field diameter~\cite{ott2016} or fiber-integrated mode matching optics~\cite{gulati2017fiber} are required. The latter uses graded index (GRIN) fiber pieces of well defined length that are spliced to the in-coupling fiber. The mirror fabrication is then performed on the last surface of the fiber stack. This approach uses the re-focusing properties of GRIN fiber to shape the fiber-sided mode. A schematic depiction is shown in Fig.~\ref{fig:fiberRealizations}~(c). 

\subsection{Tunable and stable FFPCs}
\label{sec:locking}
A key parameter in any application of FFPCs is their frequency stability and their ability to be tuned and locked to a certain frequency reference. Frequency noise in high-finesse FFPCs arises from thermal drifts, picked-up acoustic noise from the environment, electric noise in the cavity resonance tuning and from the intrinsic mechanical noise at non-zero temperature caused by vibration modes~\cite{saavedra2021tunable,janitz2017high,lee2021novel}. The first three can be reduced by appropriate thermal and acoustic isolation and by using low-noise electrical components in the experiment. How strongly ambient acoustic noise is picked-up by the device and the influence of the vibration modes onto the device stability is largely determined by the design geometry~\cite{saavedra2021tunable}. 

As FFPCs are miniaturized devices, their intrinsic vibration modes can occur at comparably high acoustic frequencies ($>\SI{1}{\kilo\hertz}$). Whilst in macroscopic cavities low frequency vibration modes can be actively damped~\cite{whittle2021approaching}, passive stability in FFPC devices is reached by pushing the lowest order vibration modes to higher frequency reducing the amount of thermal vibration noise in the system. It also allows for high locking bandwidths to actively stabilize against environmental disturbances through an electric actuation of the cavity length e.g. via piezo-electric elements~\cite{gallego2016high,janitz2017high,saavedra2021tunable}. This design paradigm is achieved through rigid device geometries, where the opposing mirrors of the FFPC are supported in close vicinity to a common base~\cite{saavedra2021tunable,janitz2017high,lee2019microelectromechanical}. 

Aside from electronic locking also other mechanisms for example through the optomechanical spring effect of an integrated membrane~\cite{rohse2020cavity} or via thermal locking~\cite{brachmann2016photothermal,gallego2016high} can be realized. The latter is based on a self locking mechanism, where thermal drifts caused by the heat induced through the intra-cavity field lead to a stabilization of the cavity resonance. Similarly, these thermal nonlinearities can even lead to deformation of the FFPC mirrors or drive the FFPC resonance into photothermal self-oscillations~\cite{konthasinghe2017self,konthasinghe2018dynamics}.

\subsection{FFPC integration and microstructured mirrors}\label{sect:functint}
To increase the functionality of FFPCs other elements such as mechanical or electrical components can be integrated into the FFPC geometry or the FFPC itself is embedded into miniaturized experiment designs e.g. chip-based ion trap experiments~\cite{lee2019microelectromechanical}. In the particular example of ion traps the advantage of small mode-volumes of FFPCs is usually not fully exploited because of uncontrolled charging of the dielectric fiber mirrors. Usually, this is mitigated by employing long cavities, however an alternative approach to reduce the effect of charging is to integrate a microstructure such as a metallic shield directly on the fiber mirror. This metallic mask is deposited on the end facet of the fiber leaving out the central part around the core. 

\begin{figure}[h]
    \centering
    \includegraphics[width=\columnwidth]{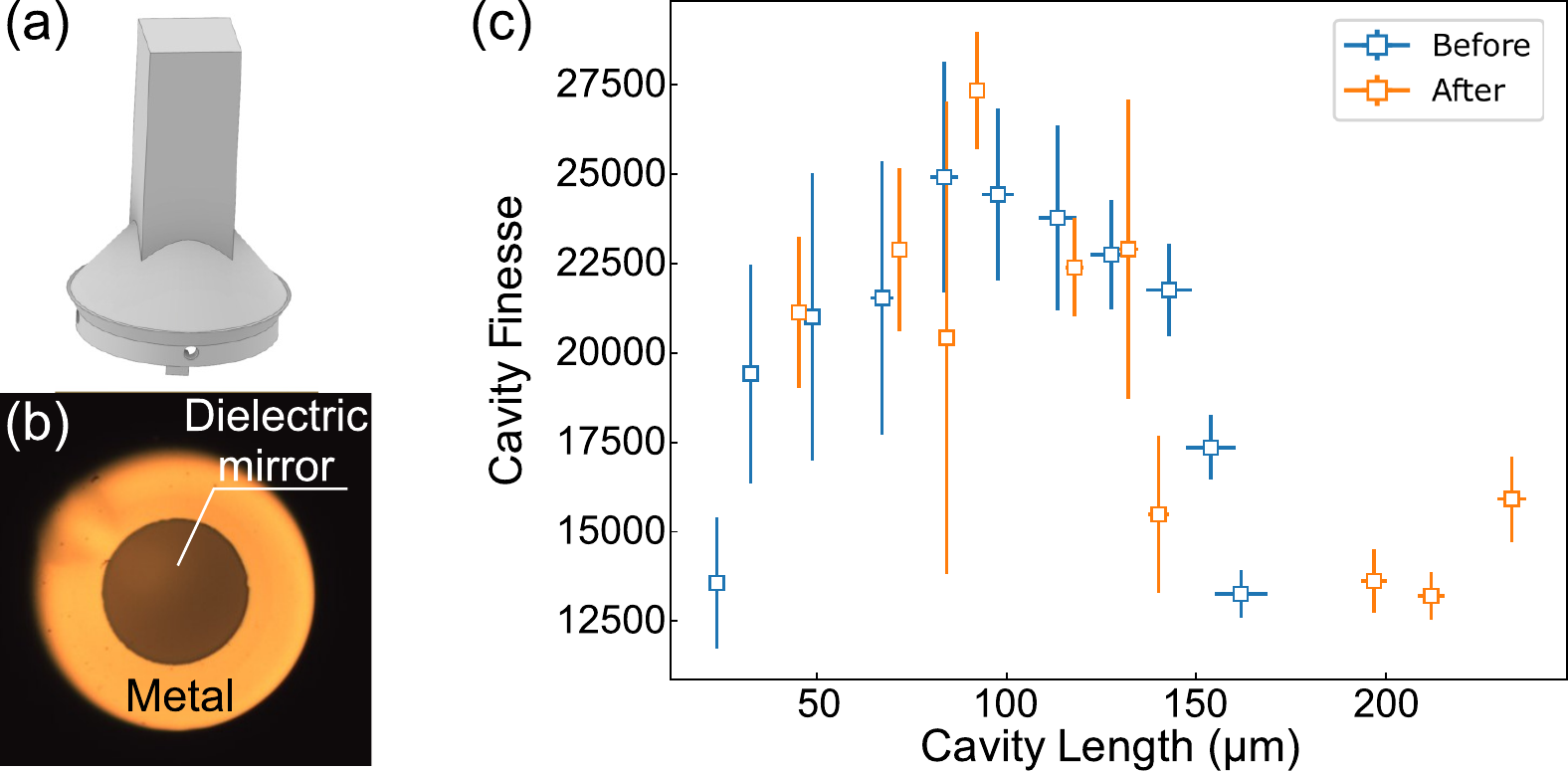}
    \caption{(a) CAD drawing of the polymer cover printed on top of the fiber end-facet. (b) Optical micrograph of the end facet of a metallized fiber. (c) Finesse of a FFPC before and after application of a metallic mask on the end-facet and the cladding of the fibers.}
    \label{fig:fiberfinesse}
\end{figure}

As an example for direct microstructuring of FFPC mirrors, a possible procedure to fabricate a metallic mask is detailed in the following. In the first step, a polymer cover (see Fig.~\ref{fig:fiberfinesse}~(a)) is printed on top of the fiber-end facet by three-dimensional laser lithography~\cite{kawata2001finer}. Subsequently, the fiber as well as the polymer cover are metallized by thermal evaporation in a deposition chamber. In order to achieve a uniform coating, the fiber is rotated during the deposition.
In the last fabrication step, the polymer cover is removed with a micromanipulator. 
Fig.~\ref{fig:fiberfinesse}~(b) exemplifies a fiber-end facet with a copper mask that we fabricated using this procedure. In order to show that the fabrication does not deteriorate the optical properties of the fiber mirrors, we characterized the FFPC before and after the application of the metallic masks. The data shown in Fig.~\ref{fig:fiberfinesse}~(c) demonstrates that the metallic mask does not have a negative effect on the cavity finesse. 

As shown here, the integration of microstructures into FFPCs can expand their capabilities, add new functions and at the same time maintain a miniaturized geometry. Future FFPC device developments are expected to more heavily make use of such additional integrated elements and embed FFPCs into other miniaturized experimental platforms.

\section{Applications}\label{sec:applications}

All applications of FFPCs make use of the miniaturized geometry leading to a small mode volume and a small footprint, see Fig.~\ref{fig:ffpcfunc}. Here we concentrate on three aspects: Small mode volumes allow to enhance the electric field strengths of a single photon to the level, where strong coupling with single atoms realizes the quantum regime (Sec.~\ref{ssec:cqed}); short cavities increase the interaction with miniaturized mechanical resonators that are easily introduced due to the open geometry (Sec.~\ref{sec:optomech}); and FFPCs also allow to construct very compact spectrometers for applications in sensing (Sec.~\ref{sec:sensing}).

\subsection{FFPCs for Cavity QED: The Quantum Limit of Light Matter Interaction}
\label{ssec:cqed}

While the radiative decay of excited atoms appears to be an unchangeable property of nature, it is not. It was realized early on by E. Purcell~\cite{purcell1946rfemission} in the context of the invention of NMR that electromagnetic cavities would allow to manipulate the coupling strength of light matter interaction by controlling the density of electromagnetic states coupling to a quantum emitter. Later on D. Kleppner~\cite{kleppner1981vacuum} coined the phrase \textit{turning off the vacuum}, which put forward the idea of \textit{cavity QED} or \textit{cQED} to investigate the quantum limit of light matter interaction, i.e. quantum emitters such as single atoms coupled to the electromagnetic vacuum or few photons. An extensive amount of information about the general status of cavity supported interaction of atoms and fields at the quantum level after more than four decades of experimenting is found in~\cite{haroche2006quantum}, and focusing on more recent applications as light matter interfaces in future quantum networks in~\cite{reiserer2010cqed}.

Today, miniaturized optical cavities integrated with optical fibers, the subject of this article, have opened yet another experimental avenue to control light matter interaction at the full quantum level.

\subsubsection{The role of the FFPC in atom-field coupling}

\begin{figure}[h]
    \centering
    \includegraphics[width=0.9\columnwidth]{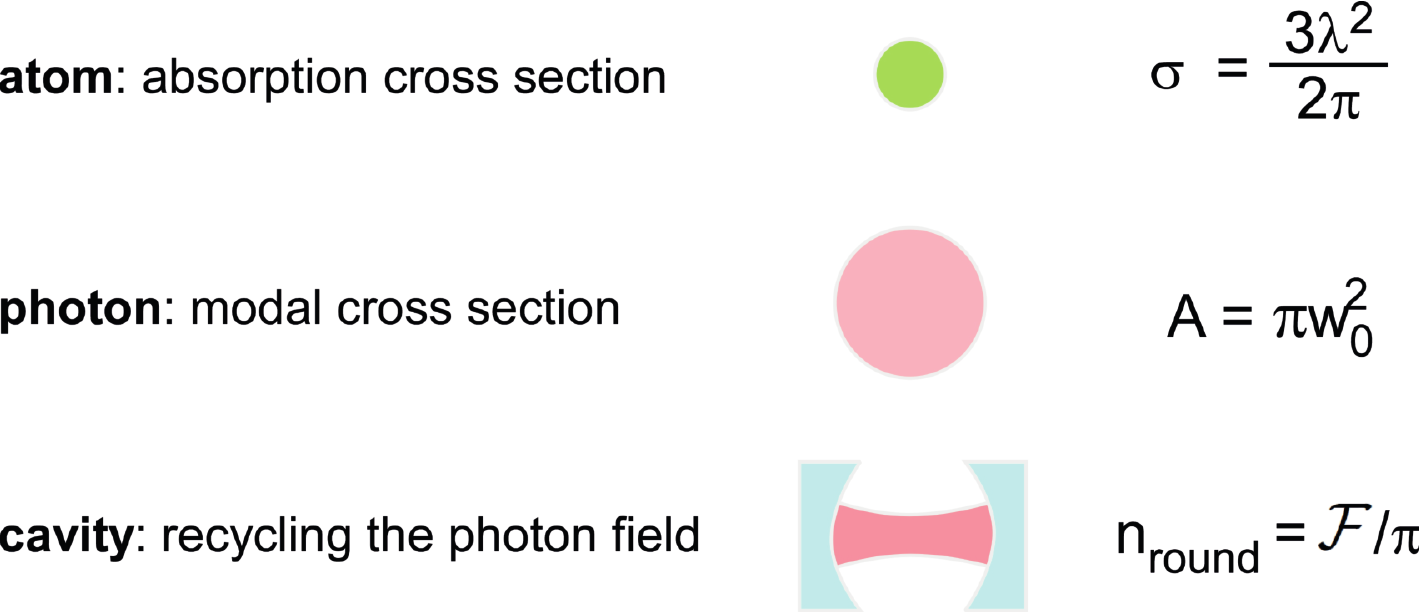}
    \caption{Simplified parameters for light matter interaction at the atom-photon level inside a cavity. The most interesting strong coupling limit occurs when the cross sections for absorption and mode at the interaction point are matched.}
    \label{fig:atomcav}
\end{figure}

FFPCs have typically small waist diameters $2w_0$ and accordingly small mode volumes Eq.\ref{eq:vmode}. In a simplified rate model, we may  characterize the interaction of light and matter by considering the absorption cross section $\sigma=3\lambda^2/2\pi$ for atomic quantum emitters at resonance wavelengths $\lambda$ and the waist area $A=\pi w_0^2$ of a photon propagating with a Gaussian beam shape (Fig.~\ref{fig:atomcav}). 

The cavity recycles the travelling photon $n_\text{round}=\cal{F}/\pi$ times with $\cal{F}$ the finesse from Eq.~\ref{eq:finesse}. The number of chances that the photon experiences to be absorbed by the atom is then estimated by the ratio of $\sigma$ and $A$ times the number of round trips, i.e.
\begin{equation}\label{eq:Cprob}
    C = \frac{\sigma}{\pi w_0^2}\cal{F}\ .
\end{equation}
One can show that this definition of the cooperativity $C$ is consistent with the more fundamental definition
\begin{equation}\label{eq:coop}
    C = \frac{g^2}{2\kappa\gamma} = \frac{1}{2}\cdot\left(\frac{g}{\gamma}\right)^2\left(\frac{\kappa}{\gamma}\right)^{-1}\ ,
\end{equation}
where the cooperativity $C$ compares the rate of evolution $g$ of the combined atom-field system coupling with the dynamic parameters governing its constituents: the free space decay rate $\gamma$ of the atomic dipole of the relevant transition ($2\gamma$ for the excitation energy), and the relaxation rate $\kappa$ of the cavity field amplitude ($2\kappa$ for the cavity field energy). Parameters in publications are typically given in terms of sets for $\{g,\kappa,\gamma\}$-values.

For an initially excited atom, the cooperativity $C$ in Eq.~(\ref{eq:coop}) determines the ratio of the emission rate into the cavity field mode with respect to the global free space emission rate by the enhancement factor
\begin{equation}\label{eq:eta}
    \eta = \frac{2C}{1+2C}.
\end{equation}
For large $\cal{F}$ though small $2C = (2\sigma/(\pi w_0^2))\cdot{\cal{F}}\ll 1$ the probability of photon emission into the cavity field is already strongly enhanced by the finesse factor $\cal{F}$ in comparison to free space. However, with $\eta \ll 1$ (Eq.~\ref{eq:eta}) the total decay rate of the atom experiences little modification, giving rise to the so-called \textit{bad cavity regime},
    \begin{equation}\label{eq:crate}
        \gamma'=\gamma(1+2C).
    \end{equation}
The realm of cavity QED has also been extended to so-called \textit{artificial atoms}, a term, which describes all quantum emitters with two and three-level quantum systems as well as good coupling to photon fields and thus resembling simple atoms. Ranging from molecules to quantum dots and more, artificial atoms offer technical advantages such as solid state samples, see Sec.~\ref{section:emitterFFPCcoupling}.

\subsubsection{FFPCs for strong atom-field coupling}
\label{subsection:strongcoupling}
\begin{figure}[h]
    \centering
    \includegraphics[width=\columnwidth]{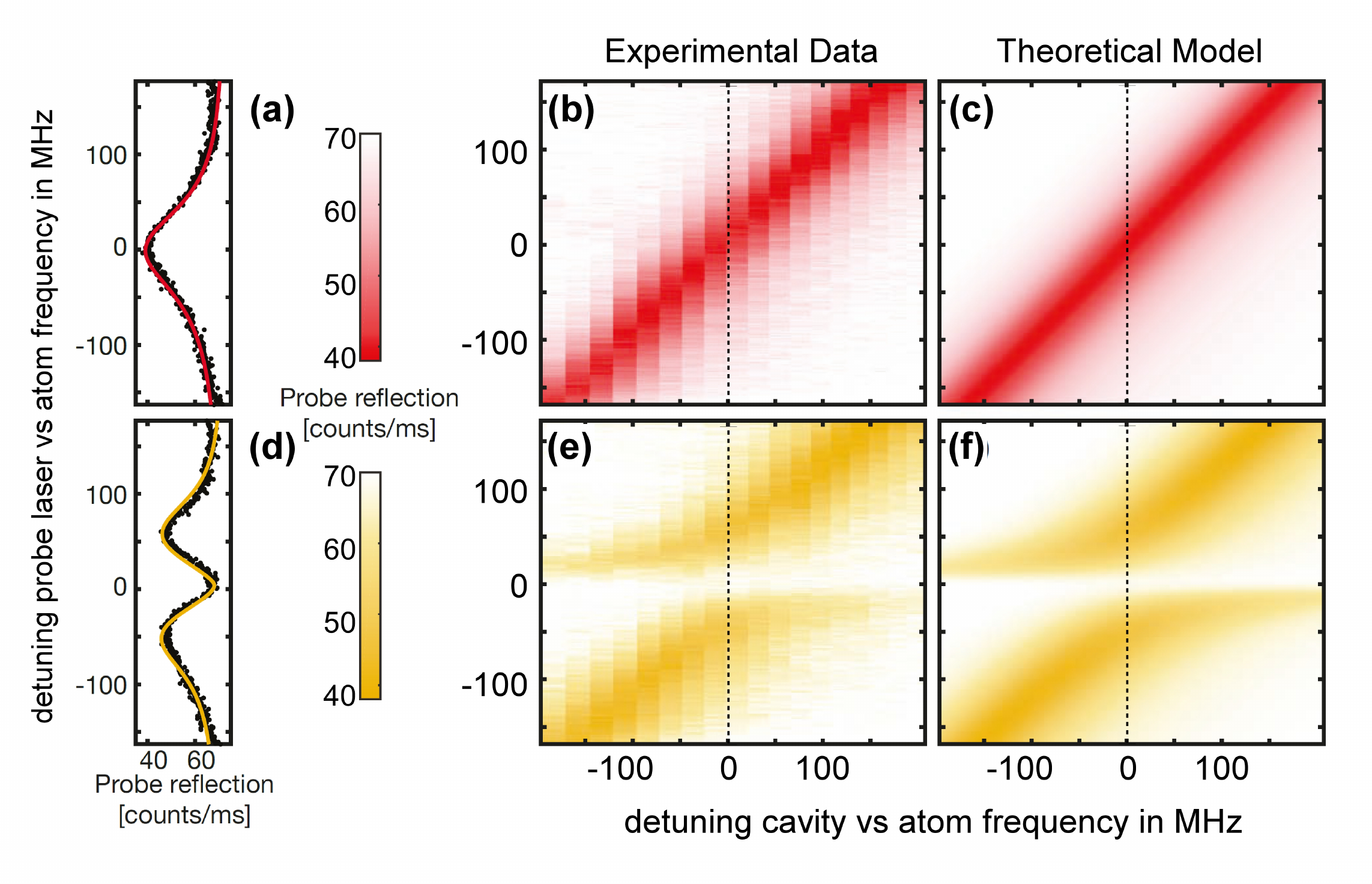}
    \caption{(a) The spectrum of the empty cavity is taken in reflection with a strong dip at resonance. (b,c) Reflection spectra for the cavity frequency set at different detunings from the atom transition frequency (dotted lines corresponding to the spectrum at resonance in (a), too). 
    (d-f) Reflection spectra with parameters as in (a-c) but with a single atom resonantly coupled to the cavity. The spectrum of the atom-cavity system shows a splitting which indicates the strong coupling regime. The gap width at resonance is given by $2g$, the atom-photon coupling rate, also called vacuum Rabi splitting.~\cite{gallego2017thesis}}
    \label{fig:split}
\end{figure}

The most interesting regime (the so-called strong coupling regime) occurs for $C > 1$ when atom-field coupling becomes so strong that the coupled systems undergo joint evolution. Coherent coupling of two quantum oscillators, atoms and fields, indeed leads to a spectrum showing a splitting as a signature of the strongly coupled system, see Fig.~\ref{fig:split}.

To achieve this regime, the small mode volumes of FFPCs offer excellent conditions: The coupling strength $g$ of atoms, or artificial atoms, with the photon field is given by the product of the atomic transition dipole moment $d$ and the local cavity field strength (Eq.~\ref{eq:uvse}). It is calculated for the energy of a single cavity photon $\hbar\omega_\text{cav}$ from  ${\cal{E}}_\text{cav}=\sqrt{2\hbar\omega_\text{cav}/\epsilon_0V_\text{mode}}$ per photon resulting in 
\begin{equation}\label{eq:gcav}
    g = \frac{1}{\hbar} d\cdot {\cal{E}}_\text{cav} = d\sqrt{\frac{2\omega_\text{cav}}{\hbar\epsilon_0 V_\text{mode}}}\ .
\end{equation}
 For $2C \gg 1$, Eq.~\ref{eq:eta} now reads $\eta \to 1$, that is, emission into free space is almost switched off and the field energy is mostly deposited into the cavity field. Since we cannot change the free space decay rate $\gamma$ of atoms, we can experimentally only control the ratio $g/\gamma$ via the field concentration caused by a small mode volume. The field relaxation rate  $2\kappa$ caused by the outcoupling rate from the resonator, however, is a function of mirror transmissivity, which is determined by the mirror coatings. Hence we can technically also choose the ratio $\kappa/\gamma$ in order to control the dynamic properties of the coupled system of atoms and fields.
 
\begin{figure}[ht]
    \centering
    \includegraphics[width=\columnwidth]{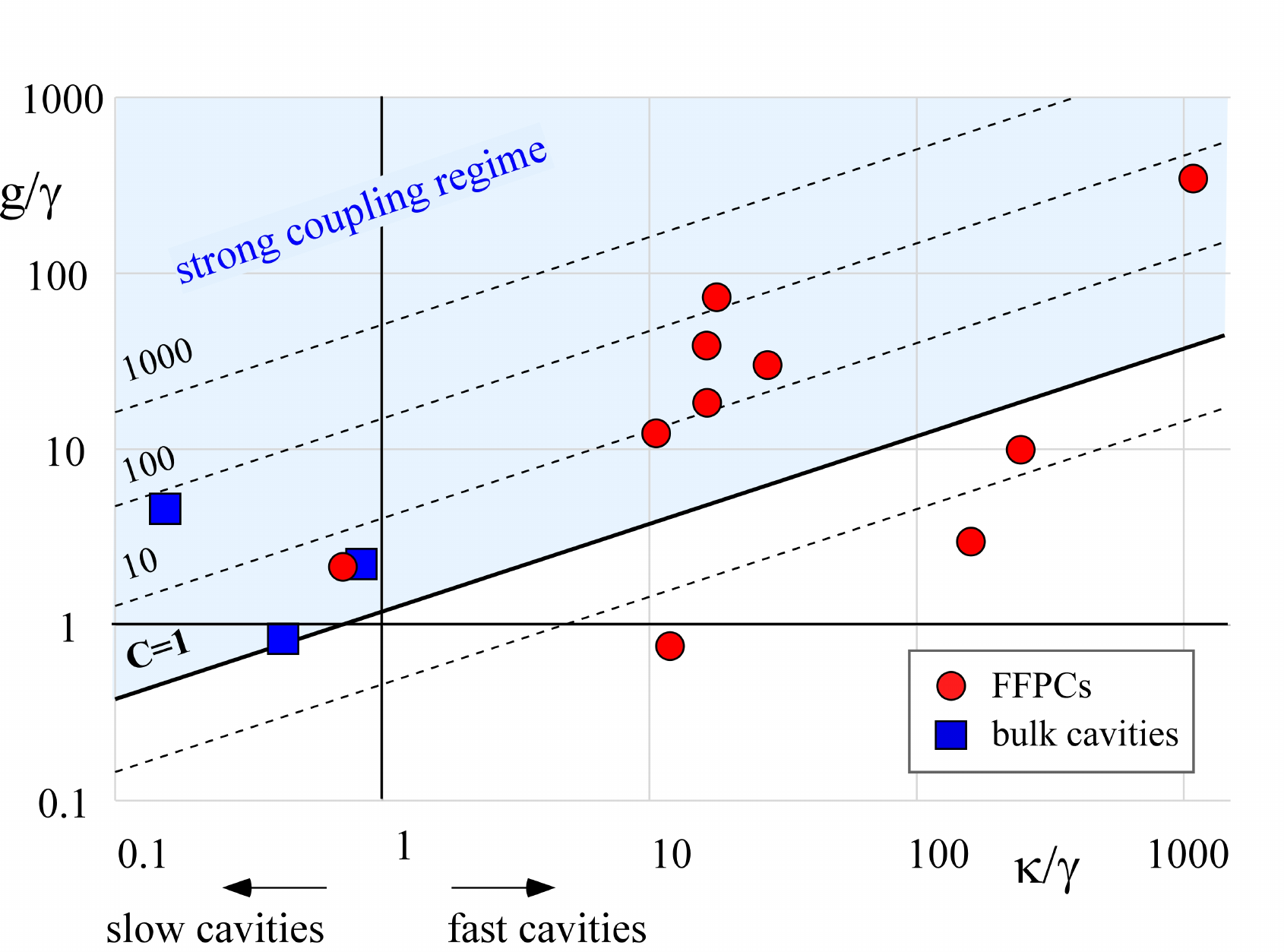}
    \caption{A non-exhaustive survey of normalized parameter sets \{$g, \kappa, \gamma$\} and $C=g^2/2\kappa\gamma$ used in cavity QED experiments with fiber based Fabry-Perot resonators (FFPCs sorted from left to right:~\cite{Takahashi2020,brekenfeld2020quantum,steiner2013,pscherer2021single,gallego2018strong,colombe2007strong,pfister2016quantum,ballance2017,lien2016}) with respect to performance in atom field coupling strength $g/\gamma$ vs. resonator field fiber coupling rate $\kappa/\gamma$. For comparison a small set of bulk cavity system parameters is shown, too, from~\cite{khuda2009,kalb2015,kuhn2002} sorted from top to bottom.}
    \label{fig:kappvsg}
\end{figure}

 In Fig.~\ref{fig:kappvsg} we compare atom-cavity systems prepared by experimenters for the strong coupling regime over the past decades with respect to their $g/\gamma$ and $\kappa/\gamma$-values for both bulk cavity experiments~\cite{khuda2009,kalb2015,kuhn2002} and FFPCs~\cite{Takahashi2020,brekenfeld2020quantum,steiner2013,pscherer2021single,gallego2018strong,colombe2007strong,pfister2016quantum,ballance2017,lien2016}. Selected sets of parameters for the ($g, \kappa, \gamma)$-values used in different laboratories show a clear tendency to take advantage of FFPC properties especially in the fast cavity regime, i.e. with $\kappa/\gamma \gg 1$.
 
 In the fast cavity domain, the energy of an initially excited atom is radiated into the cavity field at the strongly enhanced rate according to Eq.~\ref{eq:crate}. The cavity field in turn (i.e. the stored photons) is then rapidly coupled to the fiber port, too. This is experimentally advantageous since the photons leaving the cavity carry information about the quantum state of the atom-cavity system. Calling the emission of a photon from the FFPC a "read" process (since the information on the photon needed to be present in the cavity before its release), it is clear, that also the inverse "write" processes are possible and hence FFPCs coupled to quantum emitters are prime candidates for quantum memories or also certain quantum gates in future quantum networks~\cite{reiserer2010cqed}. While it is impossible to give an exhaustive survey of the applications of the different platforms of quantum emitters now coupled to FFPCs (atoms, ions, color centers, quantum dots, $\dots$), we indicate the origin of these research lines and selected current results.

\subsubsection{Quantum emitters coupling to FFPC fields}
\label{section:emitterFFPCcoupling}
As outlined above, FFPCs are favorable components to enhance the interaction of light fields with quantum emitters. Conceptually neutral atoms (and ions) are the simplest and best understood quantum light sources. With neutral atoms -- both transiting a cavity or trapped inside -- experimental progress in cavity QED was pioneered ranging from the microwave to the optical domain.

\paragraph{FFPCs for analyzing and preparing cold atom ensembles.}
The  rise of FFPCs for cavity QED was indeed triggered by experiments with a degenerate (ultra cold) system of atoms (a Bose Einstein condensate)~\cite{Steinmetz2006,colombe2007strong}: Fiber optical channels were integrated with a so-called \textit{atom chip} for analyzing and controlling properties of the atomic ensemble. The experiment showed that a system of $N$ identical atoms further increases the coupling strength (Eq.~\ref{eq:gcav}) to $g_\text{N} \propto \sqrt{N}\cdot g$. Also, quantum projection measurements (see below) allowed to create entanglement within such ultra cold atom ensembles~\cite{entangle2014haas}.

\paragraph{FFPCs for controlling single and few atom dynamics.}
With many atoms both the atom ensemble and the cavity field resemble classical harmonic oscillators since both of them exhibit the well known linear energy spectrum. The spectrum of the combined system shows of course a splitting into two modes, but the global dynamic behaviour is dominated by what is well known from two coupled classical pendulums.

This situation is very different with a single or few atoms: In the strong coupling case the atom becomes easily saturated by the field corresponding to a single photon or more. Atom-field coupling in this regime is highly non-linear with respect to the field strength that is proportional to $g_{n_\text{ph}}=\sqrt{n_\text{ph}}g$, the number of photons in the cavity. The quantum nature of this situation may be illustrated with the detection scheme for the quantum state of an atom coupled to the FFPC: from Fig.~\ref{fig:split} it is clear that an incoming light signal resonant with the empty cavity is reflected depending on the state of the cavity-atom system: For an atom in an uncoupled state, the cavity looks empty and the light field is transmitted; for a coupled state, the signal is reflected. The reflection signal thus discriminates with high efficiency an uncoupled from the coupled atomic state~\cite{qnd2011volz} which amounts to a so-called quantum non-demolition measurement.

The majority of cQED experiments with FFPCs has focused on coupling a single or few atoms (or artificial atoms, see below) to the cavity field~\cite{singleat2010gehr,gallego2018strong,brekenfeld2020quantum,macha2020nonadiabatic}. For applications, e.g. the creation of single photons on demand, the dynamics of the atom-cavity must be externally controlled. This can be accomplished by using an (artifical) atom component offering a third level which couples to the quantum states of the strongly interacting atom-cavity system. Driving the transitions involving this atomic third quantum state, the dynamics of the strongly coupled atom-field evolution can be manipulated and functional components for quantum technology such as single photon sources~\cite{gallego2018strong} and quantum memories~\cite{brekenfeld2020quantum} can be constructed. 

We finally note that trapped neutral atoms require laser cooling processes, where the fast cavity domain renders the method of cavity cooling~\cite{cavcool2013reis} inoperative and hence alternative schemes must be used~\cite{urunuela2020ground}. With respect to atom trapping the cavity may also be used to provide a dipole trap field by engineering a resonance with a suitable second wavelength~\cite{Ferri2020Mapping}.
    
\paragraph{Matching trapped ions with FFPCs.}
It is natural to consider trapped ions for cQED experiments with FFPCs. However, the small scale of the FFPCs and their dielectric nature are subject to uncontrolled charging under vacuum conditions. Therefore, ion trap experiments with FFPCs are using relatively long cavities (up to a millimeter) which increases the mode volume. Successful experiments are now carried out based on Yb$^+$~\cite{steiner2013} and Ca$^+$-ions~\cite{Takahashi2020} with wavelengths 370 and $\SI{854}{\nano\meter}$, respectively. Especially the short UV wavelength of Yb$^+$ adds to the technical challenges in maintaining high performance mirror coatings~\cite{schmitz2019ultraviolet}. At the present time investigations are underway with respect to numerous technical improvements especially with regard to ion coupled FFPCs: Unwanted stray fields are to be controlled by means of integrated electrodes (see Sec.~\ref{sect:functint}); a natural extension of FFPCs is integration with miniaturized ion traps~\cite{steiner2013,pfister2016quantum,Takahashi2020}; improved mirror fabrication for long cavities  can mitigate problems with charging of dielectric surfaces~\cite{lee2019microelectromechanical}.

\paragraph{FFPCs with color centers.}
While atoms and ions are efficient and simple quantum emitters the need for trapping devices based on laser radiation or radiofrequency fields also causes much technical overhead. Thus there are strong efforts to replace atoms by artificial atoms, which are realized with solid state host materials and cannot run away in order to simplify the systems towards applications. Prominent examples of artificial atoms -- quantum objects exhibiting atom like energy structures with transitions in the optical domain -- include color centers defects or rare earth ions in transparent materials. With nitrogen vacancies in diamond (NV centers), one of the most widespread artificial atoms, FFPCs have been shown to make NV centers at the quantum level available by e.g. demonstrating single photon light sources~\cite{albrecht2014}. A review focusing on the operation of color centers with FFPCs is found in~\cite{janitz2020cavity}. Rare-earth ion doped crystals offer another type of artificial atoms which are well known from rare earth ion doped laser crystals. They offer alternative wavelengths and coupling with FFPCs has been demonstrated to modify their radiative properties, another important step towards making photon light sources for future quantum networks at the telecom wave length around $\SI{1550}{\nano\meter}$ available~\cite{casabone2018cavity}.

\paragraph{FFPCs with molecules.}
The high spectral selectivity of FFPCs allows to also control the emission properties of molecules for selected transitions out of complex structure and at the single emitter level~\cite{toninelli2010scanning}. The enhancement of molecule-light field interaction in a FFPC based microcavity has been shown to realize nonlinear wave mixing processes at the single photon level~\cite{daqing2021,pscherer2021single}.

\paragraph{FFPCs for semiconductor quantum emitters}
Yet another natural  choice for artificial atoms is given by semiconductor nanostructures, where quantum wells and quantum dots exhibit atom like energy structures which can also be engineered over a large range of wavelengths. In contrast to color centers, quantum dots hold the promise to be pumped by simple electric currents. Within the semiconductor world the integration of all basic opto-electronic components in terms of systems is routine. However, the realization of a resonant and strongly coupled quantum dot cavity system based on quantum emitters, cavities, and waveguides remains challenging~\cite{senellart2021}. Hence the application of FFPCs offers attractive alternatives for e.g. single photon generation~\cite{trivedi2020} with respect to simplicity of fabrication or tunability. The feasibility of this approach has been shown in~\cite{miguel2013cavity,Besga2015}, where the substrate holding the quantum dots was directly integrated with DBR reflectors (DBR = Distributed Bragg Reflector). Finally, we mention that FFPCs may help to further enlarge the toolbox for processing photons with more exotic objects such as carbon nano-tubes~\cite{jeantet2017exploiting}.

\paragraph{FFPCs for hybrid systems}
Another interesting application of FFPCs with cQED concepts is concerned with the distribution of photonic quantum information by means of optical fiber links. Future quantum networks could profit from components (single-photon sources, quantum memories, $\dots$) built from different physical platforms ("hybrid systems") depending on their functional advantages. Then it is of interest to e.g. match different wavelengths as well as to control the shape of photons to allow efficient exchange between e.g. quantum dot single-photon sources and ion or atom-based quantum memories or other systems\cite{meyer2015,macha2020nonadiabatic}.

\subsection{Cavity optomechanics in FFPCs}
\label{sec:optomech}

Cavity optomechanical experiments investigate the interaction of optical cavity modes with the motion of mechanical elements~\cite{Aspelmeyer2014}. These can be mechanical resonators inside the optical cavity or, in the simplest case, one of the optical mirrors that is suspended and able to vibrate. The displacement of the mechanical resonator is thereby coupled to the number of photons in the optical mode. On one side, this leads to a push of the mechanical resonator through the radiation pressure and on the other to a detuning of the optical mode frequency $\nu$ by the mechanical resonator displacement $x$. In the simplest case the detuning $\Delta \nu$ is inversely proportional to the unperturbed length of the cavity $\ell$ as $\Delta \nu \approx \nu_0 \cdot x/\ell$~\cite{Aspelmeyer2014}. This inverse scaling with $\ell$ holds true for all optomechanical experiments, where a mechanical mode locally interacts with a distributed optical cavity mode. The most common examples for this type are suspended mirrors~\cite{groblacher2009observation} or membranes in cavities~\cite{thompson2008strong}. In other optomechanical setups, where the optical and mechanical modes are of comparable size and interacting within their full mode volume, this is typically replaced by the overlap of the two modes, as for example in the case of photoelastic coupling in optomechanical crystals~\cite{chan2012optimized}.

\begin{figure}[h]
    \centering
    \includegraphics[width=\columnwidth]{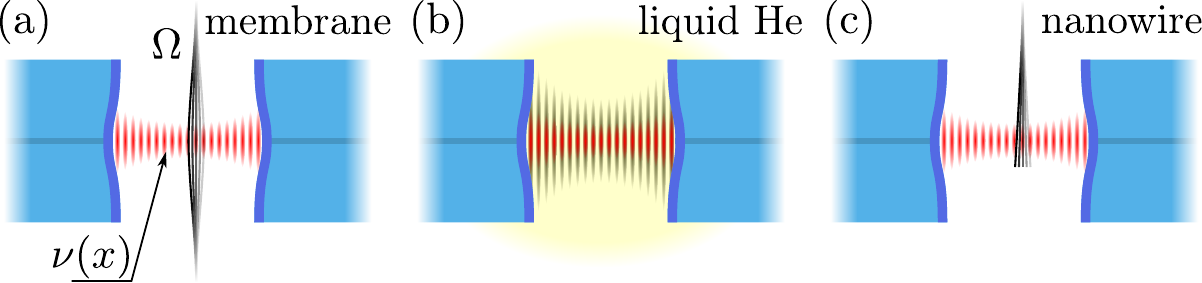}
    \caption{Cavity optomechanical experiments can be realized with a broad variety of mechanical resonators inside FFPCs. The displacement of the mechanical resonator thereby modulates the optical cavity resonance frequency e.g. by a position-dependent modulation of the effective refractive index of the optical cavity mode. (a) shows a schematic of a membrane-in-the-middle setup~\cite{thompson2008strong}. The optical field extends on both sides of the mechanical oscillator, which is clamped outside the mode volume of the optical mode. Examples for this geometry are~\cite{flowers2012fiber,rochau2021dynamical,rohse2020cavity,shkarin2014optically}. (b) depicts a setup using density waves in liquid helium as the mechanical resonance as demonstrated in~\cite{kashkanova2017optomechanics,shkarin2019quantum}. The standing wave inside the liquid helium is similarly as the optical field confined between the two mirror surfaces. (c) depicts an experiment using a nanowire as the mechanical resonator as used in~\cite{fogliano2021mapping}.}
    \label{fig:OMCschematics}
\end{figure}

FFPCs have been used in optomechanical experiments because of their ability to realize miniaturized cavities corresponding to small $\ell$ that at the same time still allow for a simple integration of mechanical elements, because of the accessible cavity volume. Most of the realizations using FFPCs can be attributed to the so-called membrane-in-the-middle (MIM) type experiments~\cite{flowers2012fiber,rochau2021dynamical,rohse2020cavity,shkarin2014optically} (see Fig.~\ref{fig:OMCschematics}~(a)), where a partially transmitting element divides the cavity into two sub-cavities. The small mode waist that can be realized in FFPCs also allows to reduce the size of the mechanical resonator in these experiments. This can be used to increase the mechanical resonator frequency $\Omega$ and decrease its effective mass $m_\text{eff}$. Pushing towards higher mechanical frequencies allows to enter the sideband resolved regime ($\Omega>\kappa$), where the cavity can be used to e.g. selectively enhance amplification or dampening of the mechanical resonator. Smaller effective masses enable larger zero point motions $x_\text{ZPM} = \sqrt{\hbar/2m_\text{eff}\Omega}$ and correspondingly a larger vacuum optomechanical coupling strength $g_0/2\pi = x_\text{ZPM}\cdot\partial\nu/\partial x$ that characterizes the interaction strength on a single-photon level. The reduction of the mechanical oscillator size has led to realizations using on-chip \chem{SiN}-~\cite{rochau2021dynamical} and nanowire oscillators~\cite{fogliano2021mapping} (see Fig.~\ref{fig:OMCschematics}~(c)), with the latter being used to map out the optical mode, position-dependent coupling and scattering. Apart from classical MIM-type experiments other realizations in FFPCs have used sound-waves of liquid helium filling the cavity volume as mechanical modes~\cite{kashkanova2017optomechanics,kashkanova2017superfluid,shkarin2019quantum} (see Fig.~\ref{fig:OMCschematics}~(b)). The distributed mechanical mode interacts with the optical mode through the modulation of the refractive index by the density variations in the acoustic standing wave pattern. 

\begin{figure}[h]
    \centering
    \includegraphics[width=\columnwidth]{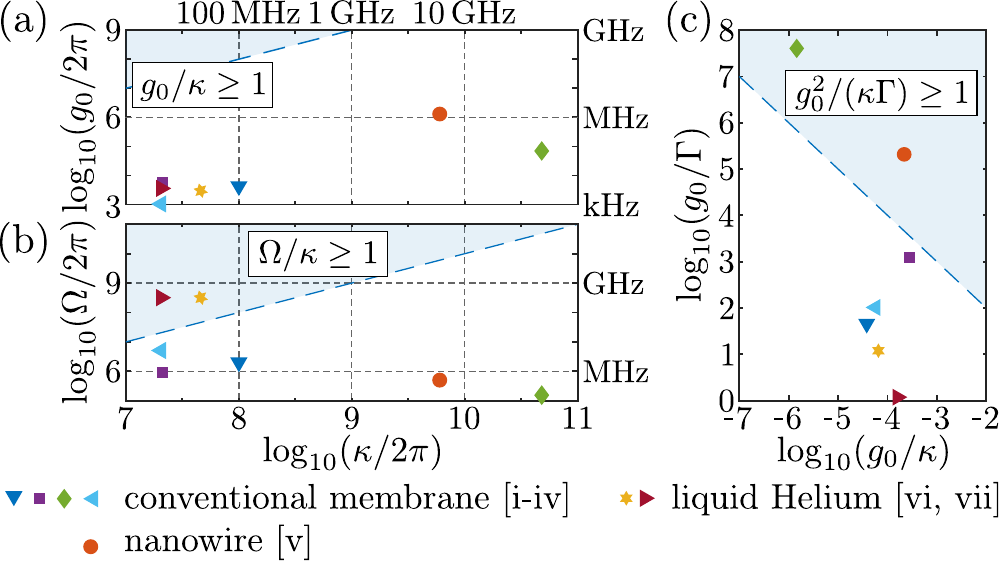}
    \caption{Comparison of optomechanical experiments using FFPCs. (a), (b) show the reached values of vacuum coupling strength $g_0$ and mechanical frequency $\Omega$ vs the cavity linewidth $\kappa$. The two sideband resolved realizations [vi, vii] use density waves in liquid helium as the mechanical resonance~\cite{kashkanova2017optomechanics,shkarin2019quantum}. The conventional membrane realizations [i-iv] are~\cite{flowers2012fiber,rochau2021dynamical,rohse2020cavity,shkarin2014optically} and the experiment using nanowire as the mechanical resonator~\cite{fogliano2021mapping}. (c) shows the performance of the different experiments in terms of the mechanical and optical loss rates $\Gamma$ and $\kappa$ in relation to $g_0$.}
    \label{fig:OMCcomparison}
\end{figure}

An overview of the reached parameters in FFPC-based optomechanical experiments is shown in Fig.~\ref{fig:OMCcomparison}. As for cavity optomechanical experiments in general, except those using cold atoms as mechanical oscillators, only moderate values of $g_0 /\kappa$ are reached. This keeps nonlinear quantum effects using single photons caused by the optomechanical interaction yet not realized. However, FFPC-based systems still reach decent parameters due to their small cavity lengths and corresponding large vacuum coupling strength and the possible high finesse. The single-photon cooperativity $C_0 = 4 g_0^2/\kappa\Gamma$ can still reach favorable values as mechanical resonators with high quality factor can be realized.  

A benefit of FFPCs in these experiments compared to even more miniaturized geometries like whispering gallery mode resonators~\cite{schliesser2010cavity} or optomechanical crystals~\cite{eichenfield2009optomechanical} lies in the tolerance to large optical drive powers. Whilst FFPCs like macroscopic optical cavities can make use of large intracavity photon numbers $n_\text{c}$ (up to Watts of optical power in high finesse cavities) to boost the linearized optomechanical coupling strength $g = \sqrt{n_\text{c}}\cdot g_0$, other platforms experience limitations caused by heating and thermal instabilities. As FFPCs can be integrated to on-chip platforms~\cite{lee2019microelectromechanical}, be used to interface very miniaturized mechanical elements, and as they are able to reach large finesse together with strong optomechanical coupling strengths~\cite{rochau2021dynamical}, they continue to be a favorable platform for cavity optomechanical experiments, especially superior where large intracavity optical powers are required.

\subsection{Cavity-enhanced sensing and other applications}
\label{sec:sensing}

\begin{figure}[h]
    \centering
    \includegraphics[width=\columnwidth]{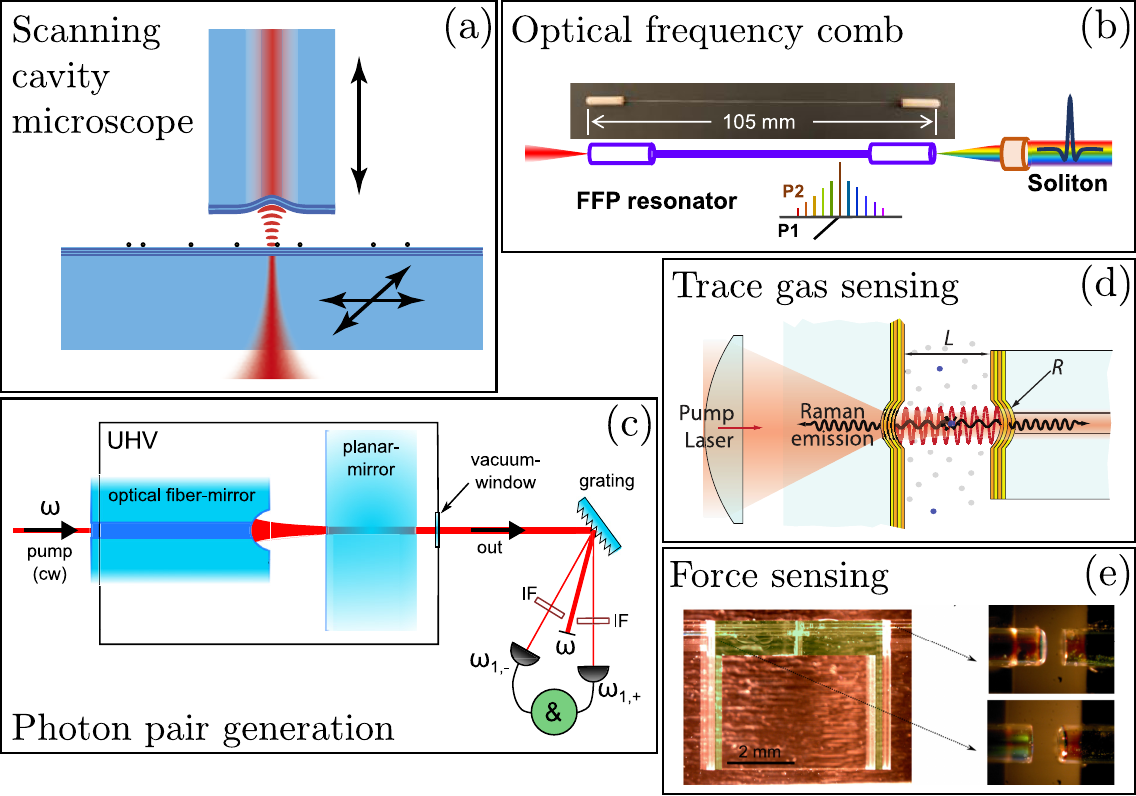}
    \caption{Compilation of various applications and experiments using FFPCs for sensing or nonlinear devices. (a) Scanning cavity microscope, adapted from Ref.~\cite{Mader2015} (CC BY 4.0). (b) Dissipative Kerr solitons and frequency comb, adapted with permission from Ref.~\cite{jia2020photonic}. Copyrighted by the American Physical Society. (c) Photon pair generation in FFPCs using the optical nonlinearity of the mirror coating, adapted with permission from Ref.~\cite{langerfeld2018correlated}. Copyrighted by the American Physical Society. (d) Trace gas sensing via enhanced Raman scattering, adapted with permission from Ref.~\cite{Petrak2014}. Copyrighted by the American Physical Society. (e) Force sensing using FFPCs, adapted with permission from Ref.~\cite{wagner2018direct}.}
    \label{fig:otherApps}
\end{figure}

Aside their utilization as an interface for quantum emitters or mechanical oscillators, FFPCs have been used for a large variety of different sensing tasks. Scanning cavity microscopes~\cite{Mader2015} (see Fig.~\ref{fig:otherApps}~(a)), use a fiber mirror as a scanning probe above a reflective substrate. They can be facilitated to map the surface~\cite{benedikter2015transverse,benedikter2019transverse} or particles located on it~\cite{Mader2015}. Measurements including multiple higher-order cavity modes are thereby used to increase the spatial resolution below the mode waist size of the fundamental mode. 

Another application of FFPCs as a cavity-enhanced sensor is for trace gas sensing~\cite{Petrak2014} (see Fig.~\ref{fig:otherApps}~(d)). There, the high intracavity intensity along with the preferred emission into cavity modes is used to enhance the emission of Raman scattering from atmospheric gases. Fiber-based sensing of substances is however not limited to gaseous media, but can also be used with liquid solutions in absorption-measurement-based~\cite{waechter2010chemical} or refractometer-based schemes~\cite{li2019fiber}.

As the resonance frequency of FFPCs is sensitive to external mechanical perturbation, it can also be used for its detection. Fiber cavity based force sensors~\cite{wagner2018direct} (see Fig.~\ref{fig:otherApps}~(e)), and fiber based sensors for strain~\cite{jiang2001simple} and vibrations~\cite{garcia2018dual} have been demonstrated.

Strong fields are a prerequisite for enhancing nonlinear optical effects. As they can be provided by optical cavities, also FFPCs have been facilitated to demonstrate and make use of optical nonlinearities. Examples include the generation of photon pairs~\cite{langerfeld2018correlated,ronchen2019correlated} (see Fig.~\ref{fig:otherApps}~e)) or in a little different cavity geometry even the realization of a photonic flywheel, which using dissipative Kerr solitons may lead to the FFPC-based frequency comb devices~\cite{jia2020photonic} (see Fig.~\ref{fig:otherApps}~(b)).

At last, also traditional interferometer applications like optical filters can be realized in fiber-based platforms. Early developments in this field even date back far before the first realization of high finesse FFPCs~\cite{stone1989optical,chraplyvy1991optical}.

\section{Conclusion and prospects}

Our summary shows that following pioneering experiments~\cite{colombe2007strong,Steinmetz2006} the improved understanding, characterization and manufacturing techniques developed for manufacturing FFPCs with excellent and robust optical and mechanical properties have opened the route for manifold applications in quantum technology, spectroscopy, and beyond. The already wide breadth of current applications  also points to a potentially even larger success in the future. FFPCs may continue to offer advantages over alternative schemes: FFPCs can overcome the low tunabilities characteristic for rigid high finesse cavities such as micro toroids etc. while the accessibility of the cavity field is unrivaled; also with respect to fully integrated photonic crystal cavities the tunability and versatility of FFPCs offer advantages. 

We expect experiments and applications using FFPCs to further diversify and demonstrate their flexibility for tackling challenges in different fields like platforms for controlled light-matter interaction, microscopy, sensing or nonlinear devices for applications at extremely low light levels. This flexibility together with their integration into fiber optics is expected to also create economic impact beside its already tremendous scientific value.

\vspace{6pt}
\begin{acknowledgements}
The authors acknowledge funding by the Deutsche Forschungsgemeinschaft (DFG, German Research Foundation, SFB/TR185 \textit{OSCAR}) under Germany's Excellence Strategy – Cluster of Excellence Matter and Light for Quantum Computing (ML4Q) EXC 2004/1 – 390534769, as well as by the Bundesministerium f\"ur Bildung und Forschung (BMBF), projects Q.Link.X and FaResQ. 
\end{acknowledgements}

\bibliography{Manuscript}
\end{document}